\documentclass{emulateapj}

\newcommand{\msol}{M_{\odot}}
\newcommand{\msola}{10^{11}~M_{\odot}}
\newcommand{\msolb}{10^{10}~M_{\odot}}

\newcommand{\redtext}{\color{black}}

\shorttitle{Strong Lens Model Comparisons}
\shortauthors{Lefor and Futamase}

\usepackage{float}
\restylefloat{table}
\usepackage{graphicx}
\usepackage[colorlinks,urlcolor=blue,citecolor=blue,linkcolor=blue]{hyperref}
\usepackage{verbatim}
\usepackage{amssymb}
\usepackage{subfigure}

\begin{document}
\title{Comparison of Strong Gravitational Lens Model Software III. Direct and indirect semi-independent lens model comparisons  of COSMOS J095930+023427, SDSS J1320+1644, SDSSJ1430+4105 and J1000+0021}

\author{Alan T.~Lefor\altaffilmark{1} and Toshifumi~Futamase\altaffilmark{1}} 
\email{alefor@astr.tohoku.ac.jp} 
\altaffiltext{1} {Astronomical Institute, Faculty of Science, Tohoku University, Sendai Japan}

\begin{abstract}
Analysis of  strong gravitational lensing data is  important in this era of precision cosmology. The objective of the present study is to directly compare the analysis of strong gravitational lens systems  using different lens model software and similarly parameterized models to understand the differences and limitations of the resulting models. The software lens model translation tool, HydraLens, was used to generate multiple models for four strong lens systems including COSMOS J095930+023427, SDSS J1320+1644, SDSSJ1430+4105 and J1000+0021. All four lens systems were modeled with PixeLens, Lenstool, glafic, and Lensmodel. The input data and parameterization of each lens model was similar for the four model programs used to highlight differences in the output results. The calculation of the Einstein radius and enclosed mass for each lens model was comparable. The results were more dissimilar if the masses of more than one lens potential were free-parameters. The image tracing algorithms of the software are different, resulting in different output image positions and  differences in time delay and magnification calculations, as well as ellipticity and position angle of the resulting lens model.  In a comparison of different  software versions using identical model input files,  results  differed significantly when using two versions of the same software. These results further support the need for future lensing studies to include multiple lens models, use of open software, availability of lens model files use in studies, and computer challenges to develop new approaches. Future studies need a standard nomenclature and specification of the software used to allow improved interpretation, reproducibility and transparency of results.

\end{abstract}


\section{Introduction}

The present time has been referred to as the "Golden Age" of Precision Cosmology. Strong gravitational lensing data is a rich source of information about the structure and dynamics of the universe, and these data are contributing significantly to this notion of precision cosmology.  Strong gravitational lens studies are highly dependent on the software used to create the models and analyze the components such as lens mass, Einstein radius, time delays etc. A comprehensive review of available software has been conducted by \cite{Lefor2012}. While many such software packages are available, most studies  utilize only a single software package for analysis.  Furthermore, most authors of strong gravitational lensing studies use their own software only. 

More recently, the status of comparison studies of strong gravitational lens models has been reviewed by \cite{Lefor2013}.  This study demonstrated that changes in redshift  affect  time delay and mass calculations in a model dependent fashion, with variable results with small changes in redshift for the same models. 

An important resource for the conduct of  comparison studies is the Orphan Lens Project, a compendium of information about strong lens systems that as of May 2014 contained data for 656 lens systems \citep{OrphanLensWeb}. There are a number of barriers to the conduct of lens model comparisons. Ideally, a comparison study of a previously studied lens would include the original model for comparison, but this is sometimes impossible because the lens model code is not made publicly available. Another barrier to performing comparative studies is the complexity of the lens model files, since there are major differences among the commonly used model software available. In order to facilitate this step of the process the HydraLens program was developed to generate model files for multiple strong gravitational lens model packages \citep{HydraLens2013}. 

To date, the largest comparison study of strong gravitational lens models was an analysis of MACSJ1206.2-0847 as part of the CLASH survey conducted by \cite{CLASHMult}. This study included four different strong gravitational lens models including Lenstool (\citep{Julio2007}), PixeLens (\citep{Saha2006}), LensPerfect (\cite{Coe2008}) and SaWLens (\cite{Merten2009}). The authors conducted five lens model analyses using the same data, and is thus categorized as a direct and semi-independent study. This type of study has great advantages in that all data and all models are available for direct comparison in a single study. 

The Hubble Space Telescope (HST) Frontier Fields project is reporting preliminary results 
\footnote{\url{http://www.stsci.edu/hst/campaigns/frontier-fields/Lensing-Models}} 
\citep{Laporte2014}.
This important deep field observing program  combines the power of the HST with gravitational lenses. Lensing analysis in the Frontier Fields project  includes models from a number of software codes including ZB, GRALE,  Lenstool, and two other non-LTM lens model software codes which  facilitate direct comparison of results from a number of lens models rather than depending on a single model from which to draw conclusions.  The Hubble Frontier Fields analysis uses models that are independently developed and optimized by each group of investigators for each code used. The power of this approach has been reported, with more results surely to follow \citep{Coe2015}. 

The goal of this study is to directly compare the results of calculations among four model software codes in the evaluation of four lens systems. The present study has several unique features. This study is the first to use computer-aided lens model design, using HydraLens software to facilitate lens model generation.There are no previous single studies which compare the results for multiple lens systems using multiple lens model software. This study was designed to further evaluate comparative lens model analyses and includes both direct and indirect semi-independent studies of four lens systems using four different software models.  Other studies have included indirect comparisons to previous lens model analyses, or direct comparisons of several lens models of a single lens system. This is the first study to also include combined indirect and direct analyses where previously published lens models were used for direct comparisons.

The nomenclature of lens model comparison studies, lens systems studied, previous lens model studies of these systems and the lens model software used are described in section \S  \ref{Methods}. The results of the lens model studies for each of the four systems studied are presented in section \S\ref{Results} and a review of existing comparison studies along with the results of this study are presented in \S \ref{Discussion}. Conclusions and suggestions for future lens model studies are in section \S \ref{Conclusions}.

\section{Methods} \label{Methods}

\subsection{Nomenclature} 
The use of standardized nomenclature to describe lensing studies is useful to evaluate multiple  studies. In this paper we follow the nomenclature previously described \citep{Lefor2013}. Lens model comparison studies are referred to as direct when the comparison is made based on calculations using two software models in the same paper, and indirect when comparison is made to previously published data. In this study, we also use the actual models from published studies (kindly supplied by the investigators) so these are considered combined indirect/direct comparisons. Lens model comparisons using the same data are referred to as semi-independent, and when different data is used, the comparison is independent. Lastly,  software is classified as Light Traces Mass (LTM, formerly known as parametric), or non Light Traces Mass (non-LTM, formerly known as non-parametric). 

\subsection{Lens Model Preparation}
Each lens model software package uses a different input data format  to describe the lens model. All of them use simple text files as input, but the format of the text files, available functionality and command structures are  dependent on the particular software. Some lens model software uses multiple accessory files to provide other data. Each of them has a unique list of commands, with great variability. HydraLens \citep{HydraLens2013} \footnote{\url{http://ascl.net/1402.023}} was written to simplify the process of creating lens model input files to facilitate direct comparison studies, and to assist those starting in the field. 

The four lens systems were evaluated using four lens model codes, necessitating 16 different models. The Lenstool model for COSMOS J095930+023427 was kindly provided by Cao \citep{Cao2013}. The glafic model for SDSS J1320+1644 was kindly provided by Rusu \citep{Rusu2013}. The remaining 14 models were written for this study using HydraLens. In the case of COSMOS J095930+023427 and SDSS J1320+1644, the two lens models we received from other investigators were used as input to HydraLens which generated the models for the other three software packages used in this study. In the case of SDSS J1430+4105 and J1000+0021, models were first written for PixelLens  \footnote{\label{Pixelens}\url{http://ascl.net/1102.007}}.  HydraLens was then used to translate the PixeLens model into the format for the other strong gravitational lens model software, including Lenstool \footnote{\label{Lenstool}\url{http://ascl.net/1102.004}}, Lensmodel \footnote{\label {Lensmodel}\url{http://ascl.net/1102.003}} and glafic \footnote{\label{glafic}\url{http://ascl.net/1010.012}}. The  translated files output from HydraLens were edited to assure that parameters were fixed or free as appropriate, and that optimization parameters were correctly set. The lens model files were then used as input to the respective lens model software.

\subsection{Gravitational Lenses Studied}
The parameters used for the four lens systems was obtained from previous studies. The geometry for each system was identical in all four models evaluated, and therefore  all  studies conducted are classified as semi-independent lens analyses. Three of the lens systems studied are listed in the Orphan Lens Database \citep{OrphanLensWeb} including COSMOS J095930+023427, SDSS J1320+1644 and SDSSJ1430+4105.

\subsubsection{COSMOS J095930+023427}
The lens COSMOS J095930 was first described by Jackson \citep{Jackson2008}. COSMOS J095930 is an early-type galaxy with four bright images of a distant background source. It is located at $z_{lens}$=0.892, and the background source is estimated at $z_{source}$=2.00. While the exact $z_{source}$ is unknown, the value used by previous investigators is 2.00. 

Models of this system were described by Faure using Lenstool \citep{Faure2011}. This model used a Singular Isothermal Ellipsoid (SIE) with external shear ($+\gamma$) and found an Einstein radius of 0.79" and $\sigma_V$=255 km s$^{-1}$.  

More recently, an extensive multi-wavelength study of this system was reported by Cao and colleagues \citep{Cao2013}, also using Lenstool.  This analysis used four different models, an SIE with two Singular Isothermal Spheres (SIS) as well as a Pseudo-Isothermal Elliptical Mass Distribution (PIEMD) model with two SIS, both with and without external shear \citep{Kassiola1993}. We  selected the SIE+SIS+SIS model used by Cao as the basis of the present indirect comparison with their work as well as the direct comparisons with the four lens models studied here.

The Lenstool model developed by Cao and coworkers was kindly supplied for this study and used as a baseline model which was then translated into input files for the other software by HydraLens. The Lenstool model used by Cao included priors for the values of ellipticity ($\epsilon =[0.0,0.9]$) and position angle (PA= [-90,90] for the SIE potential)  and for the velocity dispersion ($\sigma=[100,1000]$ for all three potentials). These same priors were used in the models of COSMOS J095930 for Lensmodel and glafic in this study. The Lenstool model developed by Cao uses optimization in the source plane. {\redtext The image positions used in all models of this system were taken from Table 1 in Cao \citep{Cao2013} }.

The Lenstool model developed by Cao has five free parameters including the velocity dispersion of the three galaxies, and orientation and ellipticity of the SIE galaxy. The positions of the second and third galaxies (SIS) in the model were fixed. The models used here were similarly parameterized. 

The present study is an indirect comparison with the analysis of Cao \citep{Cao2013} as well as a direct comparison of the four lens models studied. Since we were provided the model used by Cao, it is a combined indirect/direct comparative analysis of COSMOS J095930. All four models of this system used a $\Omega_m = 0.3$, $\Omega_{\Lambda} = 0.7$, $H_0 = 70$ km
s$^{-1}$ cosmology, as was used by \cite{Cao2013}.

\subsubsection{SDSS J1320+1644}
SDSSJ1320+1644 was initially described by \cite{Oguri2012} and \cite{Inada2013}, and is a large separation lensed quasar candidate identified in the SDSS, with a separation of $8\arcsec.585$ $\pm$0\arcsec.002 at $z_{source}$=1.487 \citep{Rusu2013}. Both an elliptical and disk-like galaxy were identified almost symmetrically between the quasars at redshift $z_{lens}$=0.899. 

A detailed lens model analysis of this system was conducted by  \cite{Rusu2013}, using glafic software. Based on their analysis, they conclude that SDSSJ1320+1644 is a probable gravitationally lensed quasar, and if it is, this would be the largest separation two-imaged lensed quasar known.  They show that the gravitational lens hypothesis implies that the galaxies are not isolated, but are embedded in a dark matter halo, using an NFW model and an SIS model. The SIS model has a $\sigma_V$=645$\pm 25$ km s$^{-1}$. We use the 'SIS free' model as the basis of the comparison study, as defined by \cite{Rusu2013}, which models the three galaxies (referred to as G1, G2 and G4) as SIS potentials and leaves the position of the dark matter halo (also modeled as a SIS) as a free parameter. The model used by Rusu includes priors for the velocity dispersion of the dark matter halo ($\sigma=[400,800]$). The same priors were used in the models of SDSS J1320+1644 in this study. The analysis by Rusu uses optimization by glafic in the image plane. {\redtext The image positions of the two images were used directly as described by Rusu \citep{Rusu2013} }.

Rusu considers models with 0 degrees of freedom, including 14 nominal constraints and the same number of nominal parameters, which fit with $\chi^2<<1$. The ellipticity and position angle are used when the position of the dark matter halo is fixed. The models developed for this study were similarly parameterized using the position of the dark matter halo as a free parameter ("SIS-free") and fixed to introduce ellipticity and position angle.  

A number of glafic models developed by Rusu and coworkers were kindly supplied for this comparative analysis and used as a baseline model which was then translated by HydraLens into models for the other software. The present study includes an indirect comparison with the analysis of Rusu \citep{Rusu2013} as well as a direct comparison of the four software lens models studied. Since we were provided a model used by Rusu, this is a combined indirect/direct comparative analysis of SDSSJ1320+1644. All four models of this system used a $\Omega_m = 0.27$, $\Omega_{\Lambda} = 0.73$, $H_0 = 70$ km s$^{-1}$ cosmology, as was used by \cite{Rusu2013}.

\subsubsection{SDSS J1430+4105}

SDSS1430+4105 was first described by \cite{Bolton2008} as part of the SLACS survey. This system is at redshift $z_{lens}$=0.285 with $z_{source}$=0.575, and has a complex morphology with several subcomponents as described by \cite{Eichner2012}. Bolton reported an effective radius of 2.55" and a $\sigma_{SDSS}$=322 km s$^{-1}$. 

 A very detailed lens model analysis of this system was then conducted by \cite{Eichner2012}. This analysis was a direct, semi-independent comparative analysis using both Gravlens (LTM) and Lensview (non-LTM) software. The authors studied five different models using Gravlens/Lensmodel, including an SIE and a Power Law (PL) model as well as three two-component de Vaucouleurs plus dark matter models. Similar results were found with the two different lens model analyses. They also studied four models using Lensview \citep{Wayth2006} including an SIE and PL models with and without external shear. We use the Gravlens/Lensmodel SIE model as the basis of the indirect comparison with their work. The plane of optimization used in the Eichner model is not explicitly stated in the report \citep{Eichner2012}. 
 
 The models developed in the previous study were not  available, and thus all models used were written for this study.   The results referred to as Model I by Eichner did not use any priors in the lens model for SDSS J1430+4105, although priors were used in the development of the model with results within the error limits reported. Similarly, priors were not used in the models in this study. The free parameters used by Eichner et al included the lensing strength b, the ellipticity and the orientation of the single-component SIE lens. These same free parameters were used in the models developed for this study. {\redtext The positions of the multiple images of this system were taken from Table 2 in \cite{Eichner2012} }. 
 
 This is both an indirect comparison (compared with the SIE model in the published study of \citep{Eichner2012}) and direct comparisons of the four lens models studied here. All four models of this system used a $\Omega_m = 0.3$, $\Omega_{\Lambda} = 0.7$, $H_0 = 70$ km
s$^{-1}$ cosmology, as was used by \cite{Eichner2012}.

\subsubsection{J1000+0021}
Using imaging data from CANDELS and the large binocular telescope, van der Wel and colleagues recently reported the quadruple galaxy-galaxy lens
J100018.47+022138.74 (J1000+0221), which is the first strong galaxy lens at $z_{lens}>$1 \citep{vanderwel2013}. This interesting system has a $z_{lens}$=1.53 and a $z_{source}$=3.417. 

\cite{vanderwel2013}, analyzed this system  in the manner described previously by \cite{vandeven2010}, and reported  an  Einstein radius of $R_E=0.35$'' with an enclosed mass of $M_E =(7.6\pm0.5)\times \msolb$  with an upper limit  on the dark matter fraction of 60\%. The highly magnified (40$\times$) source galaxy has a very small stellar mass ($\sim
  10^8~\msol$). The $z=1.53$ lens is a flattened, quiescent galaxy with a stellar mass
of $\sim 6\times\msolb$.

There have been no other lens model analyses of this system using software models and therefore all models were developed for this study using data from \cite{vanderwel2013}, and is thus is a direct comparison of the four lens software models studied. There were no priors used in the lens models of J1000+0021 in this study. The free parameter in the SIS models was only the velocity dispersion. In the SIE models, free parameters included the velocity dispersion, orientation and ellipticity. {\redtext The image positions in all models in this study for this system were taken from Table 2 in \citep{vanderwel2013}}.

All four models of this system used a $\Omega_m = 0.3$, $\Omega_{\Lambda} = 0.7$, $H_0 = 70$ km
s$^{-1}$ cosmology.


\subsection{Lens Models}
The analyses in this study were performed with four strong gravitational lens model software packages that have  been used extensively in the literature. All four systems  were modeled with all four lens model software packages. Lenstool and Lensmodel were executed under Scientific Linux version 6.4 (except as noted for Lensmodel in section \S\ref{version}), and PixeLens and glafic were executed  under OS/X version 10.9. All of these lens model software codes were reviewed in the Orphan Lens Project and the descriptions of the software are from the web site \citep{OrphanLensWeb} as well as from a review of lens model software \citep{Lefor2012}. 

Error calculations were performed according to the method of Rusu et al \citep{Rusu2013}. The errors quoted for the calculated parameters (ellipticity, orientation, magnification, time delay, etc.) reflect the calculations corresponding to calculations within the $1\sigma$ confidence interval for velocity dispersions.

The fit of the models is assessed by $\chi^2$ optimization and the RMS uncertainty. The RMS is calculated by:

\begin{equation} \label{RMS}
RMS_{images}^{2}=\sum_{i} ((x_{i}^{'}-x_{i})^2 + (y_{i}^{'}-y_{i})^2) ~/ ~N_{ima
ges},
\end{equation}
where $x_{i}^{'}$ and $y_{i}^{'}$ are the locations given by the
model, and $x_{i}$ and $y_{i}$ are the real images location, and the
sum is over all $N_{images}$ images. The $\chi^2$ results are calculated for the models by Lenstool, Lensmodel and glafic, and are reported in the data tables. The RMS value is reported by Lenstool directly, while a manual calculation was necessary for models using Lensmodel and glafic.


\subsubsection{PixeLens}
PixeLens is a non-LTM strong gravitational lens model software that is available for download \footnotemark[4] as a Java program which runs in a browser window \citep{Saha2006}. Version 2.7 was used in these studies. PixeLens is accompanied by a manual \citep{PixelensManual} and a tutorial \citep{PixelensTutorial}. PixeLens reconstructs a pixelated mass map for the lens in terms of the arrival time surface and has been used in several studies \citep{Saha2006}. 

PixeLens employs a built-in MCMC approach and creates an ensemble of 100
lens models per given image configuration. The pixelated mass map offers the advantage of being linear in the unknown. Since all equations
are linear in the unknowns, the best-fitting model and its uncertainties are obtained by  averaging over the
ensemble \citep{Saha2006, Kohlinger2013}.The pixelated mass map differentiates PixeLens from the other software used in this study which fit parametric functional forms.

\subsubsection{Lenstool}
Lenstool has been used in many different studies and is available for download \footnotemark[5] \citep{Julio2007}. Version 6.7.1 was used in these studies. Lenstool has features of both LTM and non-LTM modeling and uses a Bayesian approach to strong lens modeling and has been well-described in the literature \citep{Julio2007, Jullo2009}. There are several resources available for writing lens models for Lenstool \citep{LenstoolManual, LTDummies}. 

Lenstool can optimize most of the parameters in a model. Models produced by HydraLens for Lenstool were modified slightly to add appropriate optimization parameters and then used with Lenstool. Lenstool optimization is performed with MCMC. Lenstool uses the geometry of the images given and then finds counter-images. The image positions are recomputed and the time delays  determined.

\subsubsection{Gravlens}
The Gravlens package includes two codes, Gravlens and Lensmodel \citep{GravLens} accompanied by a user manual \citep{GravLensManual}.  Version 1.99o was used in these studies, under the Linux operating system,  downloaded from the Astrophysics Source Code Library \footnotemark[6]. However, the Darwin (Macintosh) executable file provided for version 1.99o will only run on the now obsolete PowerPC architecture. A newer version to run on the Macintosh platform under OS/X 10.9 (Gravlens version dated November 2012) was kindly provided by Professor Keeton, for these studies. Lensmodel is an extension of Gravlens and was used for all analyses here. It is fully described in two publications by Keeton \citep{GravLens, GravlensCat}, and has been used extensively. 

Lensmodel is an LTM lens model software, which optimizes the selected lens parameters and uses a tiling algorithm and a simplex method with a polar grid centered on the main galaxy. The tiles are used to determine the image positions, and then uses a recursive sub-gridding algorithm to more accurately determine image positions.

\subsubsection{glafic}
Glafic is an LTM lens model software, and includes computation of lensed images for both
point and extended sources, handling of multiple sources, a wide variety of lens
potentials and a technique for mass modeling \citep{Oguri2010} with multiple component mass models. Version 1.1.5 was used on the OS/X platform and version 1.1.6 was used with Linux in these studies\footnotemark[7]. 

Each lens is defined by the lens model and seven parameters. A large catalog of lens
models is available (including point mass, Hernquist, NFW, Einsato, Sersic, etc.).
After defining the parameters and the lens models, parameters to be varied in the $\chi^2$
minimizations are specified. Following this, the desired commands are
issued such as computing various lensing properties, Einstein radius, write lensing
properties to a FITS file, etc \citep{Glaficmanual}. Glafic has been used in a  large number of lens model studies, including SDSSJ1004 \citep{Oguri2010}, and performs lens model optimization. 

Glafic uses a downhill simplex method of optimization. The image plane is divided using square grids by an adaptive meshing algorithm. The level of adaptive meshing is set as an optional parameter.



\section{Results}\label{Results}
Each of the four lens systems  was modeled with all four lens model software codes including PixeLens, Lenstool, Lensmodel, and  glafic. Best-fit lens model parameters from previous studies are presented along with the results from this study for each system. The results reported for each lens were intended to follow the format of the data for best-fit lens parameters as reported in previous studies, and therefore there are some differences in the data presented for the four lens systems. Lenstool and glafic directly calculate the velocity dispersion and then calculate the Einstein radius and mass within the Einstein radius. Lensmodel directly calculates the Einstein radius, from which the other values were deduced. PixeLens calculates mass at various distances from the lens mass. {\redtext The figures shown are the output from each of the software packages used, and represent the graphical capabilities of that software}.

\subsection{COSMOS J095930+023427}
Best-fit lens model parameters for COSMOS J095930+023427 are shown in Table \ref{Table:J095930}. The data reported by \cite{Cao2013} are at the upper portion of the table, and show the results of the Lenstool model.  The results in this study using the Lenstool model are somewhat different because the model in this study used optimization in the image plane, rather than the source plane optimization used by Cao. The glafic model was also conducted with optimization in the image plane, while the Lensmodel model is conducted with source plane optimization because image plane optimization did not yield a satisfactory model. Direct comparisons of the four software models evaluated are shown next. The models used here were based on the SIE+SIS+SIS model used by \cite{Cao2013}. The Lenstool model includes an SIE potential at $z_{lens}$=0.892, and two SIS potentials at $z_{lens}$=0.7, as described by \cite{Cao2013}. 

The PixeLens model used image coordinates from \cite{Cao2013}, and calculated an enclosed mass inside the Einstein radius very close to that calculated by the Lenstool model. The Lenstool model optimized the ellipticity, position angle and velocity dispersion for the single SIE potential, and only the velocity dispersion for the two SIS potentials, as done by \cite{Cao2013} as free parameters. Lensmodel sets all three lens potentials at $z_{lens}$=0.892 because the software does not permit multiple lens planes.  The ellipticities and position angles optimized by each of the three codes are quite different. 

The Einstein radius of the SIE potentials are similar while there is some difference in the optimized velocity dispersions calculated by the three codes, particularly in the values calculated by glafic for the second potential. In an effort to understand this, the velocity dispersions of the first and second potentials were fixed at the values calculated by Lenstool at 234 and 412 km s$^{-1}$ respectively and the velocity dispersion of the third potential allowed to optimize, using glafic. This resulted in a velocity dispersion of 632 km s$^{-1}$ for the third potential. When the first and third values were fixed at 238 and 603 km s$^{-1}$ (as found by Cao), the second potential was optimized at 57 km s$^{-1}$. Magnifications and time delays for this model are shown in Table \ref{J095930TD}. Both time delays and magnifications calculated by all four models show great variability. 

The velocity dispersions shown in Table \ref{Table:J095930} as calculated here are slightly different from those reported by \cite{Cao2013}, because of the different optimization technique. The velocity dispersion values shown for the Lensmodel and glafic models are somewhat different. The Lenstool model used by Cao \citep{Cao2013} defined potentials at $z_{lens}$=0.892 and 0.7, although Lenstool allows only a single lens plane \citep{LenstoolManual}. When the results were re-calculated defining all lenses in the same plane ($z_{lens}=0.892$) using Lenstool, there was no effect on the calculation of the velocity dispersion. The wide variation in time delays calculated for this system are shown in Table \ref{J095930TD}, and are consistent with the wide range in time delays reported in our previous study using different models \citep{Lefor2013}. There is a wide disparity in time delay calculations seen in all of the systems evaluated in this study.

{\redtext The image plane for a representative model calculated using Lenstool is shown in Figure 1}. The image positions change from the input positions because of the image tracing algorithm used. This slight difference may account for the differences seen in time delay and magnification.  Lenstool identifies 16 total images, which are nearly superimposed at the original positions of the four images shown in Figure 1.

Each of the models uses somewhat different optimization schemes, and the velocity dispersions are a result of optimization, which may explain some of the differences shown in Table \ref{Table:J095930}. The differences in the results among the three software programs is not surprising, since this model had all three velocity dispersions as free-parameters.


\begin{deluxetable*}{lcccccccc}[!h]
\tablecolumns{9} 
\tablecaption{Best-fit lens model parameters for COSMOS J095930+023427  \label{Table:J095930}} 
\tablehead{\colhead{Software }   & 
            \colhead{RA}     & 
            \colhead{Dec}    & 
            \colhead{$\chi^2$ }   &
            \colhead{$e$}   & 
            \colhead{$\theta$}       & 
            \colhead{$R_{E}$} &  
            \colhead{$M (<R_{E})$}  &  
            \colhead{$\sigma_0$}  \\ 
            \colhead{}   & 
            \colhead{($\arcsec$)}     & 
            \colhead{($\arcsec$)}     & 
            \colhead{RMS(\arcsec)}   & 
            \colhead{}   &
            \colhead{(deg)}       & 
            \colhead{(\arcsec)} &  
            \colhead{($\msola$)}  &  
            \colhead{(km s$^{-1}$)}  } 
\startdata 
\cutinhead{Results from \cite{Cao2013}}

Lenstool &  &  &1.7    \\ 
SIE &  [0.0] &  [0.0] &  & 0.28 & -10 &  0.79 &   {$3.49 ^{+0.5}_{-0.3}$}  & 238  \\
SIS &  [-10.98] &  [0.474] &   & \nodata &  \nodata &  \nodata &  \nodata & 391 \\
SIS &  [3.52] &  [13.2] &   & \nodata &  \nodata &  \nodata &  \nodata & 603 \\

\cutinhead{Direct Comparison of Lens Models (This Study)}

PixeLens       & \nodata      & \nodata &         & \nodata & \nodata        & \nodata     & 3.51 & \nodata  \\  \\

Lenstool  &  &  & 1.2 &  \\
SIE &  [0.0] &  [0.0] &  0.06 & $0.33\pm 0.06$ & $-12\pm 8$ &  $0.79\pm0.03$ &   $3.49\pm0.7$ & $234\pm 23$  \\
SIS &  [-10.98] &  [0.474] &    & \nodata &  \nodata &  1.8 &  11.7 & $412\pm 32$ \\
SIS &  [3.52] &  [13.2] &    & \nodata &  \nodata &  4.3 &  67.9 & $632\pm 48$ \\ \\

Lensmodel   &  &  & 2.2 &   \\ 
SIE &  [0.0] &  [0.0] &  0.3  & $0.002\pm 0.002$ & $84\pm 18$ &  $0.76\pm0.06$ &   $1.91\pm0.6$ & $252\pm 23$  \\
SIS &  [-10.98] &  [0.474] &    & \nodata &  \nodata &  1.6 &  8.81 & $369\pm 36$ \\
SIS &  [3.52] &  [13.2] &    & \nodata &  \nodata &  2.3 &  18.1 & $442\pm 56$ \\ \\

glafic   &  &  & 0.9 &   \\ 
SIE &  [0.0] &  [0.0]  &   0.2 & $0.50\pm 0.08$ & $65\pm 2$ &  $0.79\pm0.02$ &   $2.06\pm0.8$ & $256\pm 23$  \\
SIS &  [-10.98] &  [0.474]  &    & \nodata &  \nodata & 0.00 &  * & $0.60\pm 0.05$ \\
SIS &  [3.52] &  [13.2]   &    & \nodata &  \nodata &  4.2 &  57.6 & $590\pm 48$ \\ \\

\enddata 
\tablecomments{Values shown in square brackets are fixed in the models. Values without brackets are the optimized/calculated values from the model. *Calculated mass at 6.10E+00$\msol$}
\end{deluxetable*}


\begin{deluxetable*}{lcccc}[H]
 \tablecolumns{5} 
\tablecaption{Magnification and Time Delays for  Four Images in COSMOS J095930+023427 \label{J095930TD}}

\tablehead{\colhead{Software }   & 
            \colhead{A}     & 
            \colhead{B}    & 
            \colhead{C}    & 
            \colhead{D} }   
            
\startdata 

PixeLens       \\
Time Delay & 0     & 0.7      & 3.4 &  0.07      \\  \\

Lenstool \\
Magnification & $7.8\pm4.3$     & $6.3\pm10$    & $8.2\pm6.2$  & $7.9\pm7.0$        \\ 
Time Delay        &  0      & $37\pm22$       & $31\pm18$ & $32\pm21$     \\  \\

Lensmodel    \\ 
Magnification & $1.5\pm1.5$      &  $1.3\pm1.5$     & $0.7\pm2.5$  & $0.9\pm1.5$     \\ 
Time Delay        &  0      & $16\pm12$       & $9.4\pm8.8$ & $9.9\pm7.5$   \\  \\

glafic    \\ 
Magnification & $-4.2\pm1.1$      & $9.2\pm2.2$      & $-102\pm7.0$ & $113\pm23$    \\ 
Time Delay        & 0      &  $28\pm7$      &  $5.0\pm1.4$ & $4.9\pm1.4$  \\  \\

\enddata 
\tablecomments{Time delay is shown in days }
\end{deluxetable*}

\begin{figure}
\centering

\includegraphics[scale=0.4]{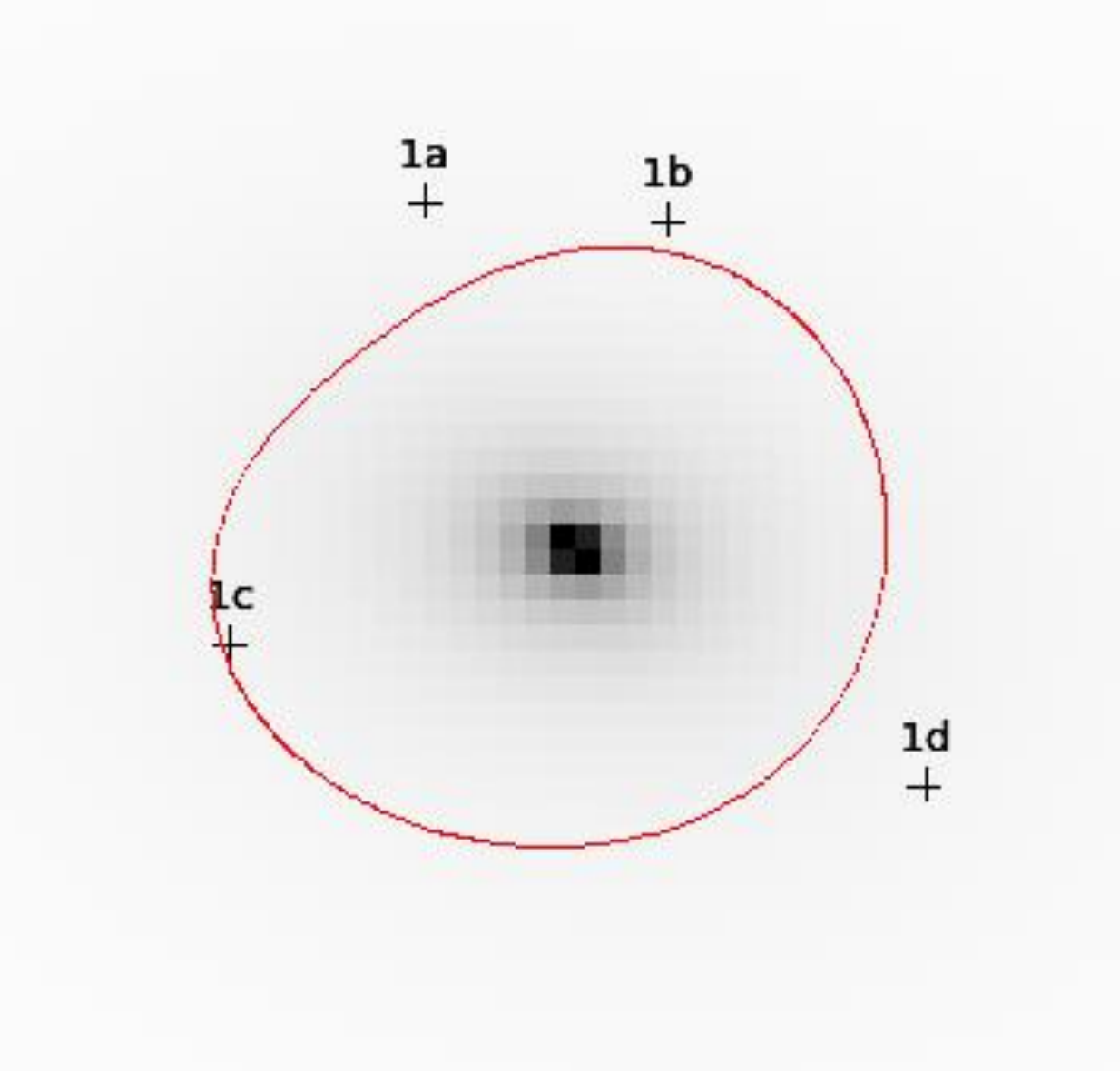}



\caption {\redtext Image plane for COSMOS J095930+023427 calculated by Lenstool. The critical line is shown in red. Each image position is shown as a cross with a label. The center of the mass distribution is shown in gray at the center}

\end{figure}



\subsection{SDSS J1320+1644}
Best-fit lens model parameters for SDSS J1320+1644 are shown in Table \ref{Table:J1320} with an indirect/direct comparison to the study of \cite{Rusu2013} and the four direct comparisons in this study. \cite{Rusu2013} utilized a glafic model that modeled the potentials of G1, G2 and G4 which were boosted by an embedding dark matter halo. One of the  published models used four SIS potentials and fixed the locations of the first three, allowing the position of the fourth (the dark matter halo) to optimize ("SIS free").  Furthermore, they concluded that any reasonable mass model reproduced the observed image configuration. The values shown in Table \ref{Table:J1320} are those as presented in the paper, as the 'SIS free' model \citep{Rusu2013}. In this study, the values calculated by \cite{Rusu2013} and shown here were reproduced exactly using their model, and the $\pm$ values are at 1$\sigma$. 

The PixeLens model has a much lower calculated time delay than the other models, and an enclosed mass within 1$\sigma$ of the value reported by \cite{Rusu2013}. As performed by \cite{Rusu2013}, the positions of the sources were kept fixed for the first three SIS potentials. The velocity dispersion and position of the last potential (the dark matter halo) were optimized. The optimized position of the fourth potential  calculated in the Lenstool model is quite different, and the velocity dispersion is similar to other models. Lensmodel uses the Einstein radius, rather than velocity dispersion so the Einstein radii for the first three SIS potentials were fixed, and the fourth was a free parameter. The mass of the fourth potential calculated by Lensmodel is nearly identical to the values calculated using glafic by \cite{Rusu2013} as well as the Lenstool and glafic models reported here. The time delays and magnification values show more variability.  

The Lenstool, glafic and Lensmodel models conducted in this study use image plane optimization, similar to the glafic analysis conducted by Rusu.  The calculated models of SDSS J1320+1644 show similar optimization for the mass of the fourth SIS potential, with fairly similar positions calculated by Lensmodel and glafic, while the positions calculated by Lenstool show greater variability. There is great variability among the calculated time delays and magnifications. 

The calculations performed in this study using glafic are the same as the glafic SIS-free model reported by \cite{Rusu2013}. Table \ref{Table:J1320} shows that the mass calculated for the fourth SIS potential, which was a free parameter, optimized to the same value for Lenstool, Lensmodel and glafic. The optimized geometry was slightly different for Lenstool compared to the others. The Einstein radius calculated by all four models was almost the same for the first SIS potential. The fact that the velocity dispersion for the fourth lens potential was optimized to the same value in all of the models may reflect the fact that there was only a single free parameter in each model. This is different from the results above with COSMOS J095930+023427, which optimized three lens potentials as free parameters, with varying results among the models tested. 

The model of SDSS J1320+1644  was  straightforward including four SIS potentials which was reproduced in all software models without difficulty.  The model used by Rusu \citep{Rusu2013} had 0 degrees of freedom and with a resulting $\chi^2<<1$, due in part to the design of the model with 14 nominal constraints and 14 parameters. The similarity of the potentials used to model the system may have contributed to the close results for optimization of the mass. Despite this, position, magnification and time delay showed great variability among the four models. The velocity dispersion for only the fourth lens potential was left as a free parameter, with the other three fixed, which is likely a major factor in the close agreement found among the various models in the calculation of the velocity dispersion. 

{\redtext The image plane of the glafic model is shown in Figure 2, which is the same as shown in Figure 6 of \cite{Rusu2013}}. The image positions in the image plane are the same as the input positions in all models. Despite this, there is variability in the time delay and magnification calculations.


\begin{deluxetable*}{lcccccccc}[!h]
\tablecolumns{9} 
\tablecaption{Best-fit lens model parameters for SDSS J1320+1644  \label{Table:J1320}} 
\tablehead{\colhead{Software }   & 
            \colhead{RA}     & 
            \colhead{Dec}    & 
             \colhead{$\chi^2$ }   &
            \colhead{$\mu$}    & 
            \colhead{$\Delta$t}       & 
            \colhead{$R_{E}$} &  
            \colhead{$M (<R_{E})$}  &  
            \colhead{$\sigma_0$}  \\ 
            \colhead{}   & 
            \colhead{($\arcsec$)}     & 
            \colhead{($\arcsec$)}     & 
            \colhead{RMS(\arcsec)}   & 
            \colhead{}  &
            \colhead{(days)}       & 
            \colhead{(\arcsec)} &  
            \colhead{($\msola$)}  &  
            \colhead{(km s$^{-1}$)}  } 
\startdata 
\cutinhead{Results from \cite{Rusu2013}}

glafic & & & $\chi^2<<1$ & $37\pm29$ & $-860\pm460$\\ 
SIS  &  [-4.991]  &  [0.117]  &   & &   &  $0.5\pm0.2$ &  $2.1\pm1.5$  &  [237] \\
SIS  &  [-2.960]  &  [3.843]  &  &  &  &   \nodata &  \nodata   &  [163] \\
SIS  &  [-9.169]  & [5.173]   &  & &  &   \nodata &  \nodata   &  \nodata \\
SIS  &   -4.687  & 1.149  &   &   & &  \nodata &   \nodata & $645\pm25$ \\

\cutinhead{Direct Comparison of Lens Models (This Study)}

PixeLens       & [0.0]      & [0.0]      &   & \nodata & 3.5        & \nodata     & 2.9 & \nodata  \\  \\

Lenstool  & & & 11.5 & $1.0 \pm 1.2$ & $ -1 \pm 10 $ \\
SIS & [-4.991]  &  [0.117] &  2.3  & &  &  [0.49] &   1.1 & [237]  \\
SIS &   [-2.960]  &  [3.843] & & &  &  [0.23] &   0.25 & [163]  \\
SIS & [-9.169]  & [5.173] & &  &  &  [0.12] &   0.07 &  [118]  \\
SIS &  -0.471 &  0.179 & &  &  &  3.6 &   61 & $645\pm 25$  \\ \\

Lensmodel  & & & 51.1 & $1.3 \pm 2.0$ & $485 \pm 210$  \\ 
SIS &  [-4.991]  &  [0.117] &  3.9  &  & &  [0.49] &   1.1 & [237]  \\
SIS &   [-2.960]  &  [3.843] & &  & &  [0.23] &   0.25 & [163]  \\
SIS &  [-9.169]  & [5.173] & &  & &  [0.12] &   0.07 & [118]  \\
SIS &  -3.93 &  2.43 & &  & &  2.9 &   53 & $576\pm 25$  \\ \\

glafic & & &2.0E-06 & $37 \pm 29$ & $-860 \pm 460$ \\ 
SIS &  [-4.991]  &  [0.117]  &  0.10  & &  &  [0.5] &   $2.1\pm1.5$ & [237]  \\
SIS &   [-2.960]  &  [3.843]  & & &  & [0.23] &   0.25 & [163] \\
SIS & [-9.169]  & [5.173] & &  & &  [0.12] &   0.070 & [118]  \\
SIS &  -4.687 &  1.149  & &  &  & $3.6\pm0.2$ &   61 & $645 \pm 25$  \\ \\

\enddata 
\tablecomments{Values shown in square brackets are fixed in the models. Values without brackets are the optimized/calculated values from the model. }
\end{deluxetable*}

\begin{figure}
\centering





\includegraphics[scale=0.4]{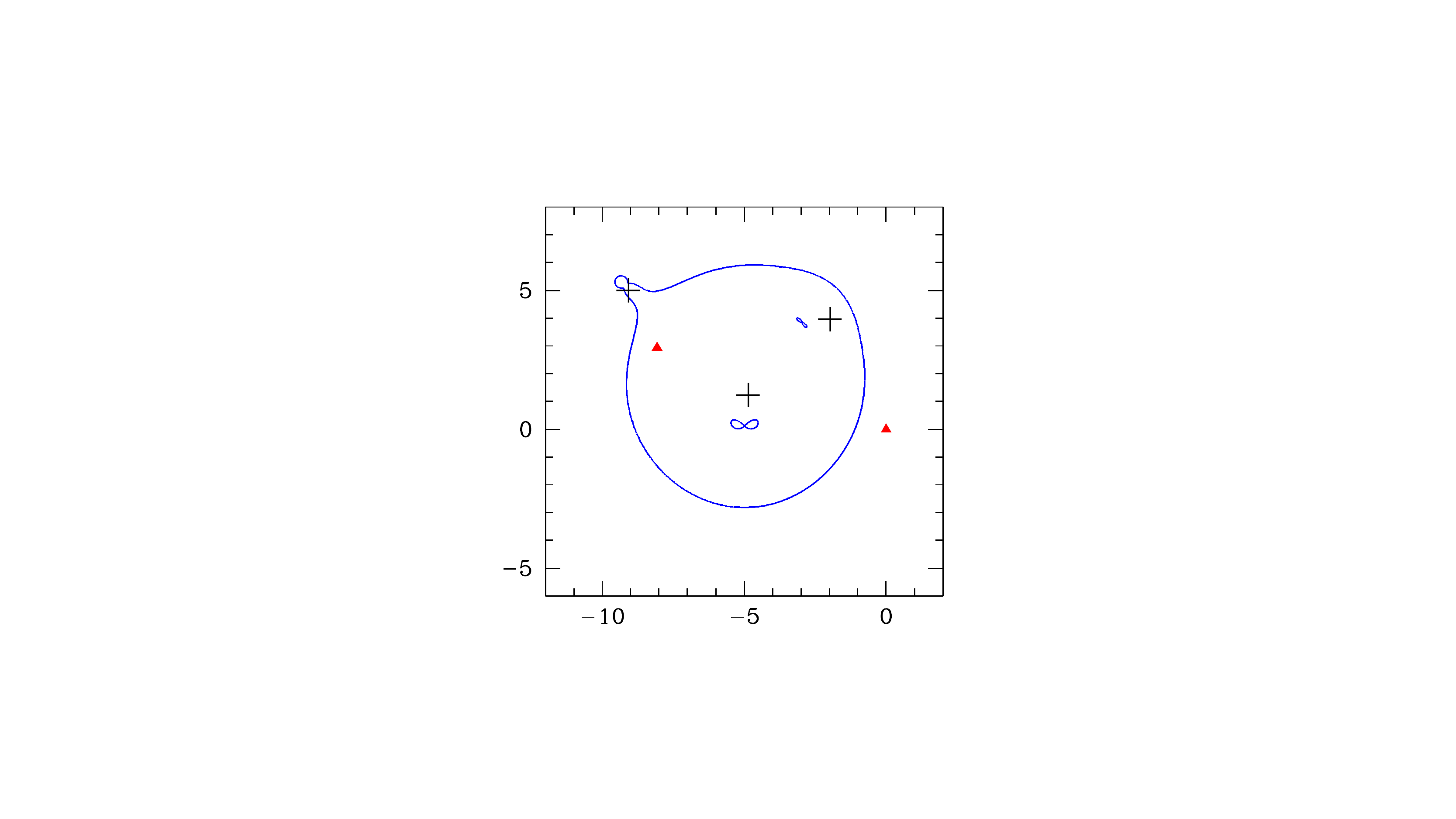}

\caption{\redtext {Image plane for SDSS J1320+1644 calculated by glafic. The blue line is the critical line. Image positions are shown as red triangles. The centers of the masses are shown as black crosses. }}

\end{figure}

\subsection{SDSSJ1430+4105}
The indirect comparison to the work of \cite{Eichner2012} and the results of the four direct comparisons in this study are shown in Table \ref{TABLE:J1430}. In \cite{Eichner2012} there are five different models tested for SDSSJ1430+4105. The models were tested with Gravlens/Lensmodel (LTM)  \citep{GravLens} and Lensview (LTM) \citep{Wayth2006}, and the results compared in a direct comparison. 

The model used in the current study is based on Model I, as described in \cite{Eichner2012}, which models the lens as an SIE, ignoring the environment of the lens. The best fitting parameters reported by \cite{Eichner2012} are shown in Table \ref{TABLE:J1430}. The results of Eichner are in good agreement with those by \cite{Bolton2008}. In the SIE model using Lensview as reported by \cite{Eichner2012}, their results were very similar to those with the Lensmodel model. The input files for the model used by \cite{Eichner2012} were not available for this study, making this study both an indirect  and direct comparison.

The Lenstool, glafic and Lensmodel models conducted in this study use image plane optimization. The enclosed mass calculated by PixeLens inside the Einstein radius, is slightly higher than the result published by \cite{Eichner2012}. The Einstein radii calculated by all the models are very close to each other as well as close to the result of \cite{Eichner2012}. As shown in other lens systems in this study, there is considerable variation in magnification and time delay calculations among the four models studied as shown in Table \ref{J1430TD}. The optimized ellipticities among the four models are all quite close, but there is significant variability in the optimal position angles calculated. 

The models used in this study (results shown  in Tables \ref{TABLE:J1430} and \ref{J1430TD}) were written without detailed knowledge of the model used by \cite{Eichner2012}. Despite this, the models all had similar results, especially in regard to Einstein radius, enclosed mass and velocity dispersion calculations. 

The image plane of the glafic model of this system is shown in Figure 3.  The glafic (Figure 3) model resulted in just 4 images in the output image plane. In contrast, Lenstool identified a total of 28 images. 

The position angles were somewhat different but there was good agreement among the models for ellipticity calculations. As with other models in this study, there was variation in the calculation of time delays and magnifications. 

One of the reasons for such close agreement among the models is that the models all used a single  SIE potential, which allowed for comparable potentials among the four lens model codes tested. There was a single lens plane in all of the models.


\begin{deluxetable*}{lcccccccc}[H]
 \tablecolumns{9} 
\tablecaption{Best-fit lens model parameters for SDSSJ1430+4105 \label{TABLE:J1430}} 
\tablehead{\colhead{Software }   & 
            \colhead{RA}     & 
            \colhead{Dec}    & 
            \colhead{$\chi^2$ }   &
            \colhead{$e$}    & 
            \colhead{$\theta$}       & 
            \colhead{$R_{E}$} &  
            \colhead{$M (<R_{E})$}  &  
            \colhead{$\sigma_0$}  \\ 
            \colhead{}   & 
            \colhead{($\arcsec$)}     & 
            \colhead{($\arcsec$)}     & 
            \colhead{RMS(\arcsec)} &
            \colhead{}    & 
            \colhead{(degrees)}       & 
            \colhead{(\arcsec)} &  
            \colhead{($\msola$)}  &  
            \colhead{(km s$^{-1}$)}  } 
\startdata 
\cutinhead{Results from \cite{Eichner2012}}

Lensmodel  & & &11.5  \\ 
SIE &  [0.0] &  [0.0] & & {$0.71 ^{+0.02}_{-0.02}$}  & {$-21.6 ^{+2.5}_{-2.3}$} &  {$1.49 ^{+0.02}_{-0.02}$} &   {$5.35 ^{+0.07}_{-0.06}$} & $322\pm22$  \\

\cutinhead{Direct Comparison of Lens Models (This Study)}

PixeLens       & [0.0]      & [0.0]      & & \nodata & \nodata        & \nodata     & 6.04 & \nodata  \\  \\

Lenstool  & & & 4.9 \\
SIE &  [0.0] &  [0.0] & 0.25 & $0.89 \pm 0.03 $ & $82 \pm 22 $&  $1.45 \pm 0.02$ &   $3.59\pm0.05$ & $317\pm22$  \\ \\

Lensmodel  & & & 15.9  \\ 
SIE & [0.0] & [0.0] & 0.30 & $0.53  \pm 0.32 $& $22 \pm 32$ &  $1.42 \pm 0.02$ &   $3.51\pm0.05$ & $309\pm22$  \\ \\

glafic   & & & 2.4 \\ 
SIE &  [0.0] &  [0.0]  & 0.29 & $0.52 \pm 0.03 $& $-10 \pm 3.4$ &  $1.50 \pm 0.02 $ &   $3.67\pm 0.11 $ & $334\pm 22 $  \\ \\

\enddata 
\tablecomments{Values shown in square brackets are fixed in the models}
\end{deluxetable*} 


\begin{deluxetable*}{lccccc}[H]
 \tablecolumns{6} 
\tablecaption{Magnification and Time Delays for  Five Images in SDSS J1430+4105 \label{J1430TD}}

\tablehead{\colhead{Software }   & 
            \colhead{A}     & 
            \colhead{B}    & 
            \colhead{C}    & 
            \colhead{D}       & 
            \colhead{E} }
            
\startdata 

PixeLens   \\
Time Delay    & 0      & 0       & 0 & 0        & 0   \\  \\

Lenstool \\
Magnification & $1.10\pm0.2$      & $2.66\pm0.1$       & $0.5\pm0.5$ & $2.7\pm0.4$        & $0.49\pm0.3$   \\ 
Time Delay        &  0      & $82\pm18$       & $66\pm22$   & $90\pm33$        & $-16\pm12$   \\  \\

Lensmodel    \\ 
Magnification & $1.0\pm0.0$      & $1.56\pm0.2$       & $0.7\pm0.15$  & $0.77\pm0.15$        & $0.36\pm0.10$   \\ 
Time Delay        &  0      & $31\pm2.5$       & $32\pm3.0$ & $32\pm3.0$       & $39\pm3.5$  \\  \\

glafic    \\ 
Magnification & $2.8\pm0.02$      & $-1.4\pm0.02$       & $-1.4\pm0.02$ & $1.1\pm0.02$        & \nodata   \\ 
Time Delay        &  0      & $34\pm4$       & $34\pm4$  & 0        & \nodata   \\  \\

\enddata 
\tablecomments{Time delay is shown in days }
\end{deluxetable*}

\begin{figure}
\centering





\includegraphics[scale=0.4]{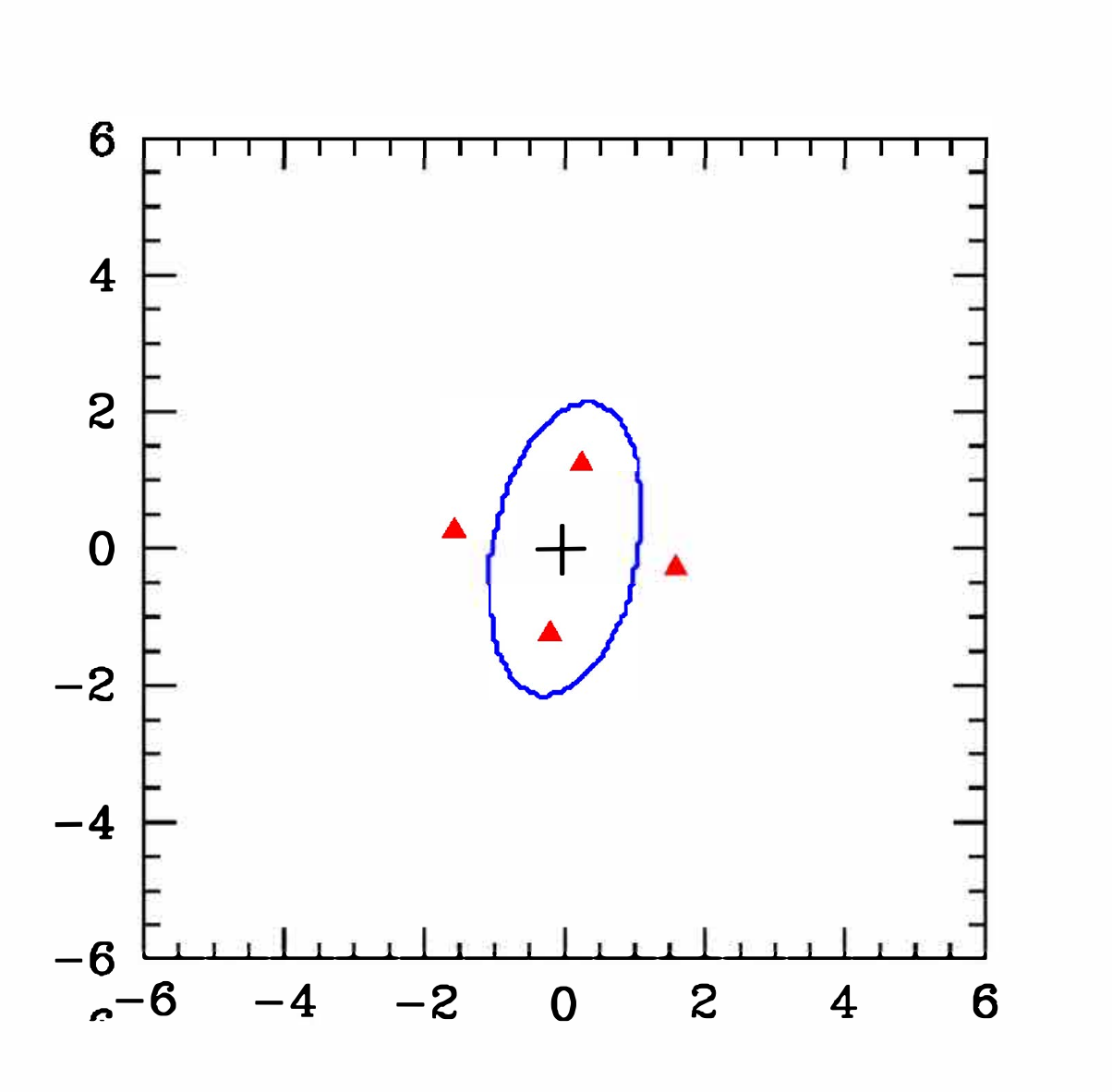}

\caption { {\redtext Image plane for SDSS J1430+4105 calculated by glafic. The blue line is the critical line. Image positions are shown as red triangles. The center of the mass density is shown as a black cross.}}

\end{figure}



\subsection{J1000+0021}
An analysis of this  lens system was performed by \cite{vanderwel2013} with a calculated  Einstein radius of
$R_E=0.35$'' (or 3.0 kpc) with an enclosed mass of $M_E =7.6\pm0.5\times \msolb$. There have been no extensive lens model analyses of this system published to date. This is the first strong galaxy lens at $z_{lens}>$1. In all models, the position (both RA and Dec) of the lens galaxy was kept constant, and the mass was a free parameter optimized by the software.  Further details of the model used were not provided, such as the model software used or the $\chi^2$ calculation.

Results of the four direct comparisons done in this study are shown in Table \ref{TABLE:J1000}. This lens system was modeled both using an SIS and an SIE, with all lens model software tested.  The Lenstool, glafic and Lensmodel models conducted in this study use image plane optimization.  The PixeLens model calculated the enclosed mass the same as reported by \cite{vanderwel2013}. Using an SIS potential, the Einstein radius, enclosed mass and velocity dispersion calculations were nearly the same for Lenstool, Lensmodel and glafic. The Einstein radii and velocity dispersions were very close to that reported by \cite{vanderwel2013}. Calculations of magnification and time delay showed quite a bit of variability in these models.

The results of the models shown in Table \ref{TABLE:J1000} show very similar results for the SIS and the SIE models. The enclosed mass within the Einstein radius is somewhat lower than that reported by \cite{vanderwel2013} for Lenstool, Lensmodel and glafic although the PixeLens model reproduced the enclosed mass calculation very well. Similar to the models used for SDSSJ1430+4105, these models were all quite straightforward with a single potential located at the origin, which may have contributed to the concordance of results. 

Comparing the results of the SIE models, the results with an SIE model using the four software packages were also nearly identical, although among the SIE models, there was some variability in the calculations of ellipticity and position angle. 

{\redtext The image plane of the glafic model of this system is shown in Figure 4.} This system is particularly interesting as the image positions in the Lensmodel and glafic models have an almost identical geometry, while the image positions in the Lenstool model are different. The time delays and magnifications in the Lensmodel and glafic models are very similar, while the Lenstool model values are  different.


\begin{deluxetable*}{lcccccccc}[H]
 \tablecolumns{9} 
\tablecaption{Best-fit lens model parameters for J1000+0021\label{TABLE:J1000}} 
\tablehead{\colhead{Software }   & 
            \colhead{RA}     & 
            \colhead{Dec}    & 
            \colhead{$\chi^2$ }   &
            \colhead{$\mu$}    & 
            \colhead{$\Delta$t}       & 
            \colhead{$R_{E}$} &  
            \colhead{$M (<R_{E})$}  &  
            \colhead{$\sigma_0$}  \\ 
            \colhead{}   & 
            \colhead{($\arcsec$)}     & 
            \colhead{($\arcsec$)}     & 
            \colhead{RMS(\arcsec)}   & 
            \colhead{}   &
            \colhead{(days)}       & 
            \colhead{(\arcsec)} &  
            \colhead{($\msola$)}  &  
            \colhead{(km s$^{-1}$)} }\\
\startdata 
\cutinhead{Results from \cite{vanderwel2013}}

 & & &  &  \\ 
  &     &     &  & $40\pm2$ &   & 0.35 &  $0.76\pm0.5$  &  $182\pm10$ \\

\cutinhead{Direct Comparison of Lens Models (This Study) - Singular Isothermal Sphere (SIS) Models} 
PixeLens       & [0.0]      & [0.0]  &     & \nodata & 2.3        & \nodata     & 0.8 & \nodata  \\  \\

Lenstool   \\
SIS &  [0.0] & [0.0] & 2.9 & $1.6\pm3.3$ & $-18\pm22$  &  $0.4\pm0.2$ &   $0.6\pm0.2$ & $200\pm10$  \\
 & & &  0.05 \\

Lensmodel    \\ 
SIS &  [0.0] & [0.0] & 5.8 & $2.3\pm0.4$ & $0.1\pm0.1$ &  $0.4\pm0.4$ &   $0.6\pm0.4$ & $190\pm10$  \\ 
 & & &  0.23 \\

glafic  \\ 
SIS &  [0.0] & [0.0]  & 0.3 & $7.1\pm12$ & $3.8\pm1.8$ &  $0.4\pm0.05$ &   $0.6\pm0.25$ & $192\pm10$  \\ 
 & & &  0.10 \\

\cutinhead{Singular Isothermal Ellipsoid (SIE) Models}

\colhead{Software }   & 
            \colhead{RA}     & 
            \colhead{Dec}    & 
            \colhead{$\chi^2$ }   &
             \colhead{$e$}    & 
            \colhead{$\theta$}       &  
            \colhead{$R_{E}$} &  
            \colhead{$M (<R_{E})$}  &  
            \colhead{$\sigma_0$}  \\ 
            \colhead{}   & 
            \colhead{($\arcsec$)}     & 
            \colhead{($\arcsec$)}     & 
            \colhead{RMS(\arcsec)} & 
            \colhead{}       & 
             \colhead{}    & 
            \colhead{(\arcsec)} &  
            \colhead{($\msola$)}  &  
            \colhead{(km s$^{-1}$)} \\
            
\tableline

Lenstool   \\

SIE & [0.0] & [0.0] & 1.7 &  $0.76\pm0.4$ & $19\pm18$ & $0.4\pm0.05$ & $0.6\pm0.07$ & $190\pm10$ \\
 & & &  0.04 \\
 
Lensmodel    \\ 

SIE & [0.0] & [0.0] &  1.7 & $0.26\pm0.04$ & $-71\pm35$ & $0.4\pm0.1$ & $0.6\pm0.05$ & $190\pm10$ \\
 & & &  0.12 \\

glafic  \\ 

SIE & [0.0] & [0.0] & 0.5 &  $0.008\pm0.003$ & $-70\pm19$ & $0.4\pm0.05$ & $0.6\pm0.22$ & $189\pm10$ \\
 & & &  0.05 \\

\enddata 
\tablecomments{Values shown in square brackets are fixed in the models}
\end{deluxetable*}

\begin{figure}
\centering





\includegraphics[scale=0.4]{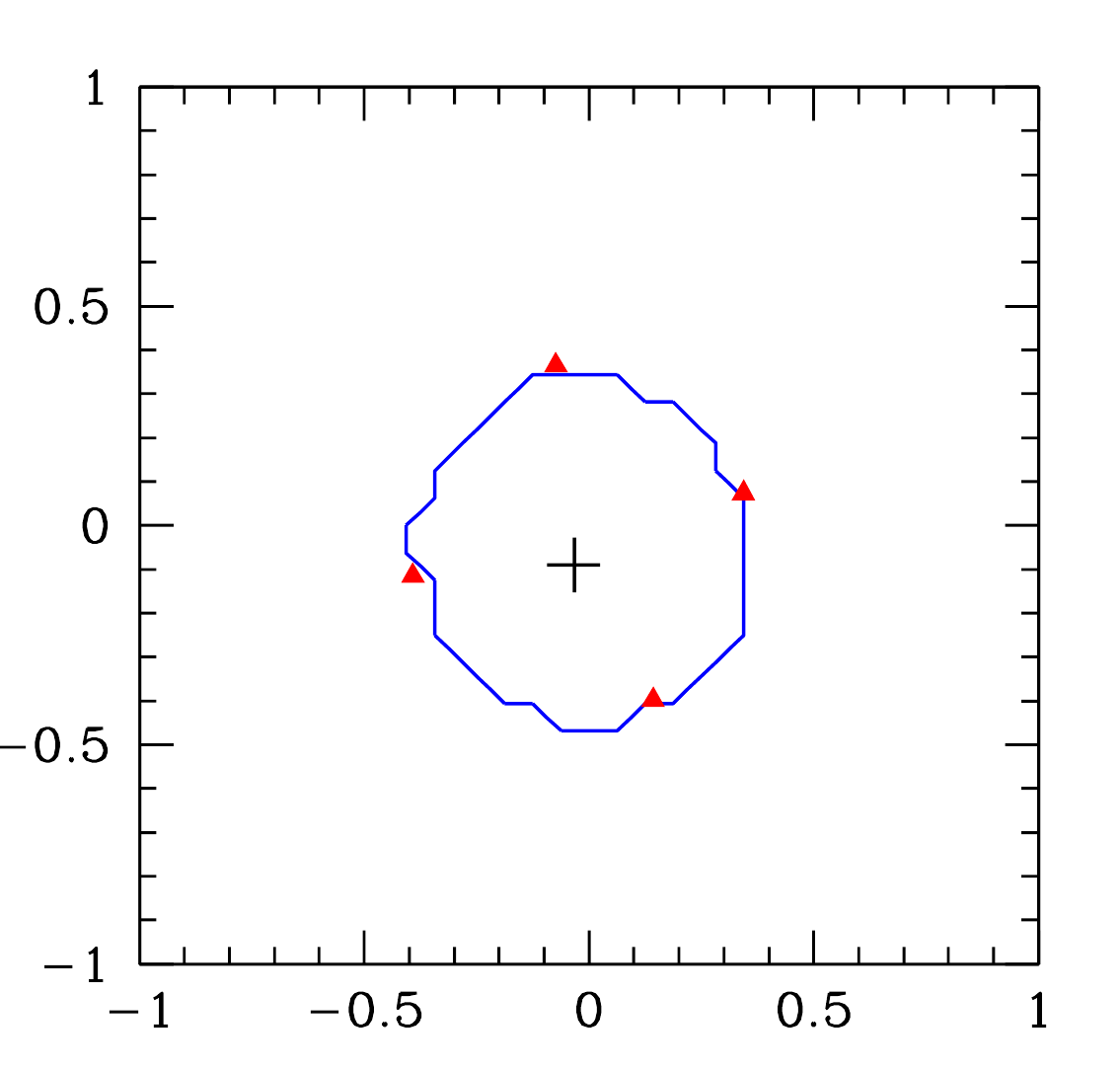}

\caption { {\redtext Image plane for J1000+0021 calculated by glafic. The blue line is the critical line. Image positions are shown as red triangles. The center of the mass density is shown as a black cross.}}

\end{figure}


\subsection{Comparison of Results}
There are some generalizations that can be made comparing the results calculated from the models for each of the four lens systems studied. The Einstein radii and mass within the Einstein radii are quite close for the four models of each system. The Einstein radius is calculated from the average distance between the lens center and multiple images, and is insensitive to the radial density profile \citep{Oguri2013}. The conversion from the Einstein radius to the enclosed mass within the Einstein radius is dependent only on the lens and source redshifts, and is therefore model independent \citep{Oguri2013}. Thus, the similar results for Einstein radii and mass within the Einstein radii are expected since all models had the same system geometry of $z_{lens}$ and $z_{source}$. 

There is variation among the calculated time delays and magnifications comparing the models generated by each of the four lens model software programs. The image positions input to each model were identical. The  image positions in the models studied change due to the ray-tracing algorithms in each software model. These differences explain some of  the variation seen in time delay and magnification. {\redtext In some cases, the use of a similarly parameterized model leads to a model that has not converged appropriately, which illustrates some of the differences in the software. This is evident in the RMS values calculated for the Lenstool and Lensmodel models of J1320+1644 }

There is also little agreement among calculations of ellipticity and position angle. The variation in results for calculated ellipticity and position angle may be a result of differences in the optimization algorithms used by Lenstool, Lensmodel and glafic. 

The complexity of the model also has an impact on agreement among the calculated values for velocity dispersion. In the models for SDSS J1430+4105, J1000+0021and SDSS J1320+1644, there was only one potential with the velocity dispersion as a free-parameter for optimization. In all three of these systems, there was close agreement among the calculated values. In the model of COSMOS J095930+023427, there were three lens potentials which were optimized, with quite a bit of variation among the results from the three software programs used.

\subsection{Comparison of Lens Model Software by Version}\label{version}
In order to evaluate the effect of software version and/or operating system / hardware platform, the model of SDSS J1320+1644 was evaluated with  glafic and Lensmodel on two different hardware platforms. Glafic is distributed as an executable file with version 1.1.5 for the OS/X platform and version 1.1.6 for Linux. Lensmodel is available as an executable file only for download as version 1.99o for the Linux platform, and we were provided a version to run on OS/X. 

Input files for the models of SDSS J1320+1644 were used unchanged. In the first test, the model was tested with the two versions of glafic. The mass of the first three SIS potentials were held as fixed parameters and the mass of the fourth potential, as well as its position, were free parameters to be optimized. Identical results were reported using either version of glafic, on both platforms. The results were identical  including the numbers of models used for optimization in each run and the calculation of all parameters evaluated. The content of all output files produced by both versions was identical. The models for SDSS J1320+1644 were then tested with each of the two versions of Lensmodel.  In this same test, optimizing the fourth SIS potential, results with Lensmodel were slightly different comparing the two versions. The optimized Einstein radius of the fourth potential using the Linux version is reported as 3.622605, and the OS/X version reports 3.622528. There are similarly small differences in the optimized position of the fourth potential. 

In the next test,  the mass of all four potentials was optimized. The results with glafic, on both hardware platforms, were again identical in regard to all parameters evaluated, to the accuracy of the last decimal place reported. The contents of all output files produced by glafic were identical with the Linux and OS/X versions. However, the two versions of Lensmodel reported widely disparate results with the two versions tested. The Einstein radii of the four optimized SIS potentials using the Linux version are 1.851, 1.004, 0.3161 and 1.660. Using the OS/X version, the four potentials are optimized at 2.234, 1.818, 0.3139 and 2.006. 

Among the various studies reported in Tables \ref{Table:Indep} and \ref{Table:Depen}, the software version used is reported in only one study. The hardware platform and/or operating system used in the calculations is not reported in any of the studies shown in these tables.


\section{Discussion}\label{Discussion}
Small changes in redshift have different effects on the calculation of time delays and mass by different lens model software codes \citep{Lefor2013}. In that study,  a mock model with a single potential and four images as well as a model of SDSS J1004+4112 were evaluated and the effect of  changes in redshift  on changes in calculations of time delay and mass  were determined. The study showed that  changes in time delay and mass calculations are not always proportional to changes in $D_{\rm d}D_{\rm s}/D_{\rm ds}$, as would be predicted.  The image positions change expectedly as a result of ray-tracing tracing algorithms which are not the same for all of the software used. This is partly responsible for the differences in the values of time delay and mass in both systems when comparing the models from four different lens model software packages.

The present study was designed to  specifically compare the results using the same models with different software, rather than changes in the results, to compare results from different codes. The present study is the largest strong gravitational lens software comparison study performed to date, evaluating four different lens systems with four different lens model software codes in a single study, and is the first study to use HydraLens for the preparation of multiple models.

\begin{deluxetable*}{lcccccc}[H]
\tabletypesize{\footnotesize}
\tablecaption{ Indirect Comparison Studies of Strong Gravitational Lens Models}
\tablehead{
\colhead{  } & \multicolumn{3}{c}{Publications}\\
\colhead{Lens System} & \colhead{Software/Reference}& \colhead{Software/Reference}& \colhead{Software/Reference}
}
\startdata
\hline \\
Abell1689**   &   LensPerfect & ZB & PixeLens  \\
    &   \cite{Coe2010} & \cite{Broadhurst2005} & \cite{Saha2006}  \\ \hline
SDSSJ1004**   &   glafic & GRALE & PixeLens   \\
   &   \cite{Oguri2010} & \cite{J1004GRALE} &  \cite{Saha2006}  \\ \hline                    
\hline \\
COSMOSJ095930   &   $*$ & Lenstool & Lenstool  \\
    &   $*$ & \cite{Cao2013} & \cite{Faure2011}  \\ \hline

SDSSJ1430   &   $*$ & Lensmodel / Lensview** &    \\
    &   $*$ & \cite{Eichner2012} &    \\ \hline
SDSSJ1320   &   $*$ & glafic &    \\
    &   $*$ & \cite{Rusu2013} &    \\ \hline
\tablenotetext{}{$*$ indicates present study with PixeLens, Lenstool, Lensmodel, and glafic, $**$ indicates that other studies of this lens have been published but are not listed. **Lensview is described in \cite{Wayth2006}}
\label{Table:Indep}
\end{deluxetable*}

\begin{deluxetable*}{lcccccc}[H]
\tabletypesize{\footnotesize}
\tablecaption{ Direct Comparison Studies of Strong Gravitational Lens Models}
\tablehead{
\colhead{Lens System} & \multicolumn{6}{c}{Software} \\
\colhead{} & \colhead{Reference}& \colhead{1}& \colhead{2}& \colhead{3}& \colhead{4}& \colhead{5}
}
\startdata
\hline \\                     
SDSSJ1430   &  \cite{Eichner2012} & Lensview** & Lensmodel &   &   &   \\
Abell1703   &   \cite{ZBGRALE} & ZB & GRALE &   &    &   \\ 
MS1358   &   \cite{ZBGRALENS} & ZB & GRALE &   &   &   \\
MACSJ1206   &  \cite{CLASHMult} & ZB & Lenstool & LensPerfect & PixeLens & SaWLens*** \\
SDSS120602   &   \cite{Lin2009} & Lensmodel & Lensview** &   &   &   \\
RXJ1347.5   &   \cite{Kohlinger2013} & glafic & PixeLens &   &   &   \\ \hline
\hline \\
J1000+0221 & $*$ & PixeLens & Lenstool & glafic & Lensmodel & \\
SDSSJ1430 & $*$ & PixeLens & Lenstool & glafic & Lensmodel & \\
SDSSJ1320 & $*$ & PixeLens & Lenstool & glafic & Lensmodel & \\
COSMOSJ095930 & $*$ & PixeLens & Lenstool & glafic & Lensmodel & \\
\tablenotetext{}{$*$ indicates the present study. **Lensview is described in \cite{Wayth2006}}
\label{Table:Depen}
\end{deluxetable*}


\subsection{Indirect Comparison Studies}
Table \ref{Table:Indep} shows a review of the existing literature where parameters have been calculated using strong gravitational lens models  and compared with other published results, and as such are referred to as "indirect comparison studies".  In the indirect comparison of COSMOSJ095930 performed by \cite{Cao2013} and \cite{Faure2011}, both analyses were conducted with Lenstool, and had very similar results for Einstein radius, mass enclosed within the Einstein radius, and other parameters. It is difficult to discern the details of the model used by \cite{Faure2011} with regard to number, type and geometry of the lens potentials used. Indirect comparisons are further complicated by a lack of available detail of the model used, making it difficult to reproduce previous results. 

\subsection{Direct Comparison Studies}
Table \ref{Table:Depen} shows previous studies where different lens models were compared in the same study, as well as the evaluations performed in the present study, all of which constitute "direct comparison studies". The direct comparisons performed of Abell 1703, MS1358, MACSJ1206 and SDSS120602 have been described in detail in \cite{Lefor2013} . The information in these direct studies was complementary in nature, leading to a greater understanding of the lens system.  The lens SDSSJ1430 was investigated  by \cite{Eichner2012} who compared the results using Lensview and Lensmodel. The Lensmodel analysis assumes point sources while Lensview uses the two-dimensional surface brightness distribution of the same system. Both analyses led to the same conclusions regarding the mass distribution of the galaxy. The two lens model techniques were indeed complementary and led to similar results.  In a comparative analysis of RX J1347.5-1145 using  glafic and PixeLens, the authors note a 13 percent difference in the calculation of mass enclosed within the Einstein radius \citep{Kohlinger2013}. They suggest that the LTM model used by glafic may not be assigning sufficient mass to the profiles in the models used. We observed a similar underestimation of enclosed mass by non-LTM models as compared to PixeLens in the analysis of J1000+0021.

\section{Conclusions}\label{Conclusions}
Indirect comparison studies are of value, but as some of the comparisons conducted in this study show, it may be difficult to reproduce the results of previous studies without  previous model files available  to create the models for other software,  thus limiting the nature of the comparisons performed.  In the analyses of COSMOS J095930+023427 and SDSS J1320+1644, being able to use the same models as used in the original studies, qualifies these as direct comparisons. This supports the importance of sharing lens model files in future studies. 

Even in direct comparisons, the results with one model may not be exactly the same as with another because of the difficulty in translating some of the features of one model to another because of the differences in features of the available software. For example, it is not possible to parameterize a PixeLens model exactly the same as a Lenstool model because of inherent differences in the software. These differences may explain the observations of \citep{Kohlinger2013} as well as some of the results in this study. Despite best efforts to similarly parameterize two models, there still may be small differences. This suggests that using several models to understand a system may lead to improved understanding.

In seeking agreement among various models, the number of free parameters for the lens potentials is an important factor. While there was reasonable agreement among the calculated values for Einstein radius in single potential models, such as SDSS J1430+4105 and J1000+0021 in this study, there was less agreement in a more complicated model such as COSMOS J095930+023427, which may be a reflection of using more lens potentials to describe the system. 

Differences noted in time delay and magnification calculations may be due to differences in the image tracing algorithms used by each of the software models. The input image positions are the same in all models.  The software  calculates new positions  based on the software specific ray-tracing algorithm used going from the source plane back to the image plane, resulting in differences in time delay results. The differences in optimization algorithms used also leads to some of the observed differences among the software models, with great variation in the calculation of  ellipticity and position angle. 

These results demonstrate that there are significant differences in results using lens models prepared with different software, and are consistent with a previous study of differences in lens models  \citep{Lefor2013}. There is no intention to suggest that a particular group of models are necessarily more correct, but only to suggest that future lensing studies should evaluate lens models using several approaches to understand the system more thoroughly, as already being conducted in the Hubble Frontier Fields project. 

Based on the results of this study, in order to allow comparisons across studies, it will be important to use a consistent nomenclature for lensing studies, specifying indirect vs. direct comparisons, independent vs. semi-independent comparisons and the type of model being used as LTM vs. non-LTM, as we have previously described \citep{Lefor2013}. Furthermore, this study has shown at least in one situation that the software version used can significantly affect the results which stresses the importance of specifying the software version number being used in all future studies, in addition to the hardware/operating system platform. It is also suggested that more detail is provided in future studies to allow reproducibility of the models such as the number and types of potentials used along with the name of the potential used in the various software packages. 

One of the most important aspects of any scientific experiment is reproducibility. In gravitational lens model studies, this is impossible in many cases because the software is not  available to other investigators, or the lens model files are not available. Code-sharing of software in astrophysics is essential, as emphasized by \cite{Shamir2013}. Based on the studies reported here, the sharing of lens model files in gravitational lens studies is also essential to assure reproducibility and increased transparency in future gravitational lensing studies. Another approach in lensing studies that has been successfully applied in weak lensing is computer challenges. The use of multiple approaches including comparative studies of lens models, open software, open lens model files, and computer challenges will help to assure increased transparency in future studies and enhance the results.

\acknowledgments 
The contributions of  Rusu and Oguri (J1320 model, glafic) and Shuo and Zhang (J095930 model, Lenstool) are greatly appreciated. Their willingness to freely share their models contributed significantly to this work. Thanks is also expressed to Professor C. Keeton for providing the latest version of Gravlens/Lensmodel. We gratefully acknowledge the careful consideration of the anonymous reviewers which afforded us the opportunity to improve and clarify this manuscript. This research was supported by a Grant-in-Aid for Scientific Research from the JSPS (Grant Number 26400264).

\end{document}